\documentstyle[12pt]{article}
\newcommand{\nc}{\newcommand}
\nc{\al}{\alpha}
\nc{\g}{\gamma}
\nc{\G}{\Gamma}
\nc{\D}{\Delta}
\nc{\la}{\lambda}
\nc{\La}{\Lambda}
\nc{\var}{\varphi}
\nc{\ba}{\beta_\al}
\nc{\bb}{\beta_\beta}
\nc{\ga}{\g^\al}
\nc{\gb}{\g^\beta}
\nc{\kvt}{\sqrt{t}}
\nc{\hn}{h^\vee}
\nc{\kn}{k^\vee}
\nc{\dab}{{\delta_\al}^\beta}
\nc{\pa}{\partial}
\nc{\nn}{\nonumber \\ }
\nc{\hf}{\frac{1}{2}}         
\nc{\paj}{P_{-\al}^j}
\nc{\vmab}{V_{-\al}^\beta}
\nc{\vab}{V_\al^\beta}
\nc{\vib}{V_i^\beta}
\nc{\db}{\pa_\beta}
\nc{\dtb}{\delta_\theta^\beta}
\nc{\fabc}{{f_{ab}}^c}
\nc{\rton}{R_2^{(n)}}
\nc{\rjn}{R^j_{(n)}}
\nc{\binomial}[2]{\left (\begin{array}{c} {#1}\\ {#2} \end{array}
\right )}
\nc{\ben}{\begin{equation}}
\nc{\een}{\end{equation}}
\nc{\bea}{\begin{eqnarray}}
\nc{\eea}{\end{eqnarray}}
\nc{\cpp}{{C_+}^+}
\nc{\cnp}{{C_0}^+}
\nc{\cmp}{{C_-}^+}
\nc{\cmn}{{C_-}^0}
\nc{\cmm}{{C_-}^-}
\nc{\bra}[1]{\langle {#1}|}
\nc{\ket}[1]{|{#1}\rangle}
\nc{\C}{\mbox{\hspace{1.24mm}\rule{0.2mm}{2.5mm}\hspace{-2.7mm} C}}
\nc{\Nat}{\mbox{\hspace{.04mm}\rule{0.2mm}{2.8mm}\hspace{-1.5mm} N}}
\nc{\spa}{\hspace{1 cm},\hspace{1 cm}}
\nc{\vs}{\vspace}

\nc{\NP}[1]{Nucl.\ Phys.\ {\bf #1}}
\nc{\PL}[1]{Phys.\ Lett.\ {\bf #1}}
\nc{\CMP}[1]{Commun.\ Math.\ Phys.\ {\bf #1}}
\nc{\PR}[1]{Phys.\ Rev.\ {\bf #1}}
\nc{\PRL}[1]{Phys.\ Rev.\ Lett.\ {\bf #1}}
\nc{\PTP}[1]{Prog.\ Theor.\ Phys.\ {\bf #1}}
\nc{\PTPS}[1]{Prog.\ Theor.\ Phys.\ Suppl.\ {\bf #1}}
\nc{\MPL}[1]{Mod.\ Phys.\ Lett.\ {\bf #1}}
\nc{\IJMP}[1]{Int.\ Jour.\ Mod.\ Phys.\ {\bf #1}}
\nc{\IM}[1]{Invent.\ Math.\ {\bf #1}}
\nc{\SJNP}[1]{Sov. J. Nucl. Phys.\ {\bf #1}}

\begin{document}

\topmargin -5mm
\oddsidemargin 5mm

\begin{titlepage}
\setcounter{page}{0}
\begin{flushright}
NBI-HE-97-12\\
AS-ITP-97-10\\
April 1997\\
\end{flushright}

\vs{8mm}
\begin{center}
{\Large FREE FIELD REALIZATIONS}\\[.2cm]
{\Large OF 2D CURRENT ALGEBRAS,}\\[.2cm]
{\Large SCREENING CURRENTS AND PRIMARY FIELDS}

\vs{8mm}
{\large Jens Lyng Petersen}\footnote{e-mail address:
 jenslyng@nbi.dk},
{\large J{\o}rgen Rasmussen}\footnote{e-mail address: 
jrasmussen@nbi.dk},\\[.2cm]
{\em  The Niels Bohr Institute, Blegdamsvej 17, DK-2100 Copenhagen \O,
Denmark}\\[.5cm]
{\large Ming Yu}\footnote{e-mail address:
yum@itp.ac.cn}\\[.2cm]
{\em  Institute of Theoretical Physics, Academia Sinica, P.O.Box 2735, 
Beijing 100080, Peoples Republic of China}

\end{center}

\vs{8mm}
\centerline{{\bf{Abstract}}}
In this paper we consider Wakimoto free field realizations of simple
affine Lie algebras, a subject already much studied. We present three new 
sets of results. (i) Based on quantizing differential operator realizations
of the corresponding Lie algebras we provide general universal very simple 
expressions for all currents, more compact than has been 
established so far. 
(ii) We supplement the treatment of screening currents of the first kind, 
known in the literature, by providing a direct proof of the properties 
for screening currents of the second kind.
Finally (iii) we work out explicit free field realizations of primary fields 
with general non-integer weights. We use a formalism where the 
(generally infinite) multiplet is replaced by a generating function primary 
operator. These results taken together allow setting up integral 
representations
for correlators of primary fields corresponding to non-integrable degenerate
(in particular admissible) representations.\\[.4cm]

\end{titlepage}
\newpage
\renewcommand{\thefootnote}{\arabic{footnote}}
\setcounter{footnote}{0}

\section{Introduction}

Since the work by Wakimoto \cite{Wak} on free field realizations of affine
$SL(2)$ current algebra much effort has been made in obtaining similar 
constructions in the general case, a problem in principle solved by 
Feigin and Frenkel \cite{FF} and further studied by many groups, 
\cite{GMMOS,BF,BMP,KOS,Ito0,Ito,Dot,Ku,ATY,Tay,deBF}. Free field realizations
enables one in principle to build integral representations for correlators in
conformal field theory \cite{DF,FZ,FGPP,An}. In a recent series of papers we
have carried out such a study for affine $SL(2)$ 
\cite{PRY1,PRY2}. It turns out that screening operators of both the first and  
the second
kinds are crucial for being able to treat the general case of degenerate 
representations \cite{KK} and admissible representations \cite{KW}. It is also
necessary to be able to handle fractional powers of free fields. We have 
established well defined rules for that \cite{PRY1,PRY2}. We were particularly
interested in this technique because of its close relationship with 
2-dimensional quantum gravity and string theory \cite{HY,AGSY}, although
many other applications may be envisaged, see e.g. \cite{Sch}.

In this paper we provide the ingredients for generalizations to affine
algebras based on any simple Lie algebra. That would enable one e.g. to treat 
the case of W-matter coupled to W-gravity.

Our new results consist first in presenting
very explicit universal compact expressions for the affine currents.
We use techniques based on ``triangular" parameters on representation
spaces to treat in an efficient way any representation.  
Some of these expressions are new. Second, we have provided a proof of the 
properties of the screening currents of the second
kind proposed without proof by Ito \cite{Ito0} in addition to the better known 
ones of the first kind. Our proof for the
validity of this second kind so far works only for $SL(N)$ but it
seems natural to expect the 
result to hold in general (\cite{JR2}). Finally, we have generalized the very 
compact form of the primary field used in \cite{FGPP,PRY1,PRY2} to the general
case. A number of these results were given in preliminary form in \cite{JR}.
Primary fields for integrable representations are described in Ref. 
\cite{FF}. Our treatment holds also for non-integrable, degenerate
(including admissible) representations. The compactness of our result
is due to the use of triangular parameters.

The paper is organized as follows.
In Section 2 we fix our notation which we keep rather general. We define our
``triangular" coordinates and we introduce a crucial matrix depending on them 
in the adjoint representation of the underlying algebra. All our explicit 
results are given in a very simple way in terms of that matrix.

In Section 3 we present differential operator realizations of simple Lie
algebras. This technique is well known. The new aspect is that we
work out in great detail certain Gauss decompositions
of relevant group elements. These are the key to our explicit formulas.
We then discuss differential operators later
to become essential counterparts of the screening currents of both kinds. 
We provide several non-trivial polynomial identities later to be used.

In Section 4 we quantize the differential operator realization of a simple 
Lie algebra to a Wakimoto free field realization of the corresponding
affine Lie algebra in the standard way. The non-trivial part is to take
care of multiple contractions (or in other words the normal ordering) by
adding anomalous terms to the lowering operators. These terms were recently
discussed in the general case by de Boer and Feh\'er \cite{deBF}. Our result
again is somewhat more explicit. We end the section by listing further 
polynomial identities following from the quantum realization, to be used later
on.

In Section 5 we discuss the screening currents. First, we review the known
results for screening currents of the first kind and for completeness write
them down in our notation and indicate an explicit straightforward proof.
The new result concerns our proof of the properties of
screening currents of the second kind, generalizing the idea of 
\cite{BO} from $SL(2)$ to any simple algebra \cite{Ito0}. 
In the case of $SL(N)$ we prove that our explicit expression fulfils
the required properties.

In Section 6 we give a thorough discussion of primary fields using the 
formalism based on the ``triangular" parameters. We derive simple and
general free field realizations of primary fields with arbitrary, possibly 
non-integral weights i.e. non-integer Dynkin labels and non-integer level.

Section 7 contains concluding remarks.

\section{Notation}

Let {\bf g} be a simple Lie algebra of dim {\bf g} = $d$ and rank {\bf g} = 
$r$.
{\bf h} is a Cartan subalgebra of {\bf g}. The set of (positive) roots
is denoted ($\Delta_+$) $\Delta$, and we write $\al>0$ if $\al\in\Delta_+$.
The simple roots are $\{\al_i\}_{i=1,...,r}$. $\theta$ is the highest root,
while $\al^\vee = 2\al/\al^2$ is the root dual to $\al$. 
Using the triangular decomposition 
\ben
 \mbox{{\bf g}}=\mbox{{\bf g}}_-\oplus\mbox{{\bf h}}\oplus\mbox{{\bf g}}_+
\een
the raising and lowering operators are denoted $e_\al\in$ {\bf g}$_+$ and
$f_\al\in$ {\bf g}$_-$ respectively with $\al\in\Delta_+$, and 
$h_i\in$ {\bf h} are the Cartan operators. 
We let $j_a$ denote an arbitrary Lie algebra element. 
For simple roots we sometimes write $e_i=e_{\al_i}, f_i=f_{\al_i}$.
The 3$r$ generators $e_i,h_i,f_i$ are the Chevalley generators.
Their commutator relations are
\bea
\left[h_i,h_j\right]=0&&\left[e_i,f_j\right]=\delta_{ij}h_j\nn
  \left[h_i,e_j\right]=A_{ij}e_j&&\left[h_i,f_j\right]=-A_{ij}f_j
\eea
where $A_{ij}$ is the Cartan matrix. 
In the Cartan-Weyl basis we have
\ben
 [h_i,e_\al]=(\al_i^\vee,\al)e_\al\spa [h_i,f_\al]=
-(\al_i^\vee,\al)f_\al
\een
and
\ben
 \left[e_\al,f_\al\right]=h_\al=G^{ij}(\al_i^\vee,\al^\vee)h_j
\een
where the metric $G_{ij}$ is related to the Cartan matrix as
$A_{ij}=\al_i^\vee\cdot\al_j=(\al_i^\vee,\al_j)=
G_{ij}\al_j^2/2$, while the
Cartan-Killing form (denoted by $\kappa$ and $tr$) is
\ben
 tr(j_aj_b)=\kappa_{ab}\spa
  \kappa_{\al,-\beta}=\kappa(e_\al f_{\beta})=\frac{2}{\al^2}
  \delta_{\al,\beta}\spa \kappa_{ij}=\kappa(h_i h_j)=G_{ij}
\een
The Weyl vector $\rho=\hf\sum_{\al>0}\al$ satisfies
$\rho\cdot\al_i^\vee=1$.
We use the convention ${f_{-\al,-\beta}}^{-\g}=-{f_{\al\beta}}^\g$.
The Dynkin labels $\La_k$ of the weight $\La$ are defined by
\ben
 \La=\La_k\La^{(k)}\spa \La_k=(\al_k^\vee,\Lambda)
\een
where $\left\{\La^{(k)}\right\}_{k=1,...,r}$ is the set of fundamental
weights satisfying
\ben
 (\al_i^\vee,\La^{(k)})=\delta_i^k
\een 
Elements in $\mbox{\bf g}_+$ (or $\mbox{\bf g}_-$) or vectors in 
representation spaces (see below) are parametrized using ``triangular 
coordinates" denoted by $x^\al$, one for each positive root. We introduce the
Lie algebra elements
\ben
 e(x)=x^\al e_\al \in \mbox{\bf g}_+ \spa f(x)=x^\al f_\al \in \mbox{\bf g}_-
\een
and the corresponding group elements $g_+(x)$ and $g_-(x)$
by
\ben
 g_+(x)=e^{e(x)} \spa g_-(x)=e^{f(x)}
\een
Also we introduce the matrix representation, $C(x)$, of $e(x)$ in the adjoint 
representation
\ben
 C_a^b(x)={C(x)_a}^b={(x^\beta C_\beta)_a}^b=-x^\beta {f_{\beta a}}^b
\label{cadj}
\een
and use the following notation for the (block) matrix elements
\ben
 C=\left(\begin{array}{lll}\cpp & 0 & 0\\
                 \cnp & 0 & 0\\
                 \cmp & \cmn & \cmm
         \end{array}  \right)
\label{C}
\een
${C_+}^+$ etc are matrices themselves. In ${C_+}^+$ both row and column 
indices are positive roots, in $\cmn$ the row index is a negative root and
the column index is a Cartan algebra index, etc.
One easily sees that (leaving out the argument $x$ for simplicity)
\bea
{(C^n)_+}^+&=&({C_+}^+)^n\nn 
{(C^n)_0}^+&=&{C_0}^+({C_+}^+)^{n-1}\nn
{(C^n)_-}^0&=&({C_-}^-)^{n-1}{C_-}^0\nn
{(C^n)_-}^-&=&({C_-}^-)^n\nn  
 {(C^n)_-}^+&=&\sum_{l=0}^{n-1}(\cmm)^l\cmp(\cpp)^{n-l-1}
  +\sum_{l=0}^{n-2}(\cmm)^l\cmn\cnp(\cpp)^{n-l-2}\nn
0&=& {(C^n)_+}^0= {(C^n)_+}^-= {(C^n)_0}^0= {(C^n)_0}^- 
\eea
We shall use repeatedly that $C_\al^\beta(x)$ vanishes unless 
$\al <\beta$, corresponding to ${C_+}^+$ being upper triangular with 
zeros in the diagonal. Similarly, ${C_-}^-$ is lower triangular. 
It will turn out that we shall be able to provide remarkably simple universal
analytic expressions for most of the objects we consider using the 
matrix $C(x)$. This will be one of the new results in this paper.

For the associated affine algebra, the operator product expansion, OPE, 
of the associated currents is
\ben
 J_a(z)J_b(w)=\frac{\kappa_{ab}k}{(z-w)^2}+\frac{\fabc J_c(w)}{z-w}
\label{JaJb}
\een
where regular terms have been omitted. $k$ is the central extension and
$\kn=\frac{2k}{\theta^2}$ is the level. In the mode expansion
\ben
 J_a(z)=\sum_{n=-\infty}^\infty J_{a,n}z^{-n-1}
\een
we use the identification
\ben
 J_{a,0}\equiv j_a
\een
The Sugawara energy momentum tensor is
\bea
T(z)&=&\frac{1}{\theta^2(\kn+\hn)}\kappa^{ab}:J_aJ_b:(z)\nn
    &=&\frac{1}{t}:\sum_{\al>0}\frac{1}{\al^2}(E_\al F_\al+F_\al E_\al)
       +\frac{1}{2}(H,H):(z)
\eea
where we have introduced the parameter
\ben
 t=\frac{\theta^2}{2}\left(\kn+\hn\right)
\een
and where $\hn$ is the dual Coxeter number. This tensor has central charge
\ben
 c=\frac{\kn d}{\kn+\hn}
\label{c}
\een

The standard free field construction 
\cite{Wak,FF,GMMOS,BMP,KOS,Ito0,Ito,Dot,Ku,ATY}
consists in introducing for every positive 
root $\al>0$, a pair of free bosonic ghost
fields ($\ba,\ga$) of conformal weights (1,0) satisfying the OPE
\ben
 \ba(z)\gb(w)=\frac{\dab}{z-w}
\een
The corresponding energy-momentum tensor is
\ben
 T_{\beta\g}=:\pa\ga\ba:
\label{Tbg}
\een
with central charge
\ben
 c_{\beta\g}=d-r
\een
We will understand ``properly" repeated root indices as in (\ref{Tbg})
to be summed over the positive roots.

For every Cartan index $i=1,...,r$ one introduces a free scalar boson $\var_i$
with contraction
\ben
 \var_i(z)\var_j(w)=G_{ij}\ln(z-w)
\een
The energy-momentum tensor 
\ben
 T_\var=\hf:\pa\var\cdot\pa\var:-\frac{1}{\kvt}\rho\cdot\pa^2\var
\een
has central charge
\ben
 c_\var=r-\frac{\hn d}{\kn+\hn}
\een
This follows from Freudenthal-de Vries strange formula 
$\rho^2=\hn\theta^2d/24$.
The total free field realization of the Sugawara energy-momentum tensor 
is $T=T_{\beta\g}+T_\var$ as is well known (see also \cite{deBF}).

The vertex operator
\bea
 V_\Lambda(z)&=&:e^{\frac{1}{\kvt}\Lambda\cdot\var(z)}: \nn
\Lambda\cdot\var(z)&=&\Lambda_i\var_j(z)G^{ij}
\eea
has conformal weight
\ben
 \Delta(V_\Lambda)=\frac{1}{2t}(\Lambda,\Lambda+2\rho)
\een
It is also affine primary corresponding to highest weight $\Lambda$. A 
new result in this paper will be the explicit general construction
of the full multiplet of primary fields, parametrized by the $x^\al$ 
coordinates in Section 6.

\section{Differential operator realizations}

Following an old idea (see e.g. \cite{Kos}), elaborated on in
\cite{FF,BMP,Ito0,Ito,ATY},
we here discuss a differential operator 
realization of a simple Lie algebra
{\bf g} on the polynomial ring $\C[x^\al]$.
We introduce the lowest weight vector in the (dual) representation space
\ben
 \langle\Lambda|f_\al=0\spa\langle\Lambda|h_i=\La_i\langle\Lambda|
\een
An arbitrary vector in this representation space is parametrized as
\ben
\bra{\Lambda,x}=\bra{\Lambda}g_+(x)
\een
The differential operator realizations $\tilde{J}_a(x,\pa,\Lambda)$ 
with $\pa_\al=\pa_{x^\al}$ denoting partial derivative wrt $x^\alpha$, 
are then defined by
\ben
\bra{\Lambda,x}j_a=\tilde{J}_a(x,\pa,\Lambda)\bra{\Lambda,x}
\een
Obviously these satisfy the Lie algebra commutation relations.
It is convenient to have a similar notation for highest weight (ket-) vectors
\bea
\ket{\La,x}&=&g_-(x)\ket{\La} \nn
j_a\ket{\La,x}&=&-J_a(x,\pa,\La)\ket{\La,x}
\label{ketdef}
\eea
where the relation between the two sets of realizations of the Lie algebra,
$\{\tilde{J}_a(x,\pa,\La)\}$ and $\{J_a(x,\pa,\La)\}$, is as follows
\bea
 E_\al(x,\pa,\La)&=&-\tilde{F}_\al(x,\pa,\La)\nn
 F_\al(x,\pa,\La)&=&-\tilde{E}_\al(x,\pa,\La)\nn
 H_i(x,\pa,\La)&=&-\tilde{H}_i(x,\pa,\La)
\label{tilde}
\eea

We write the Gauss decomposition of $\langle\Lambda|g_+(x)e^{tj_a}$ 
for $t$ small as
\bea
 \langle\Lambda|g_+(x)\exp(te_\al)&=&\langle\Lambda|\exp\left( 
  x^\g e_\g+t\vab(x)e_\beta+{\cal O}(t^2)\right)\nn
 &=&\langle\Lambda|\exp\left(t\vab(x)\db+{\cal O}(t^2)\right)g_+(x)\nn
 \langle\Lambda|g_+(x)\exp(th_i)&=&\langle\Lambda|\exp\left(th_i\right)
  \exp\left(x^\g e_\g+tV_i^\beta(x)e_\beta+{\cal O}(t^2)\right)\nn
 &=&\langle\Lambda|\exp\left(t\left(V_i^\beta(x)\db+\Lambda_i\right)
  +{\cal O}(t^2)\right)g_+(x)\nn 
 \langle\Lambda|g_+(x)\exp(tf_\al)&=&\langle\Lambda|\exp\left(
  tQ_{-\al}^{-\beta}(x)f_\beta+{\cal O}(t^2)\right)\exp\left(
  t\paj(x)h_j+{\cal O}(t^2)\right)\nn
 &\cdot&\exp\left(x^\g e_\g+t\vmab(x)e_\beta+
  {\cal O}(t^2)\right)\nn
 &=&\langle\Lambda|
  \exp\left(t\left(\paj(x)\Lambda_j+\vmab(x)\db\right)
  +{\cal O}(t^2)\right)g_+(x)
\label{Gauss1}
\eea
It follows that the differential operator realization is of the form
\bea
\tilde{ E}_\al(x,\pa)&=&\vab(x)\db\nn
\tilde{ H}_i(x,\pa,\Lambda)&=&\vib(x)\db+\Lambda_i\nn
\tilde{ F}_\al(x,\pa,\Lambda)&=&\vmab(x)\db+\paj(x)\Lambda_j
\label{defVP}
\eea
Since $\tilde{E}_\al(x,\pa,\Lambda)=\tilde{E}_\al(x,\pa)$ is independent of 
$\Lambda$ it may be defined through a Gauss decomposition alone.

{}From the realization of $\tilde{E}_\al$ we obtain
\ben
 \vab(x) tr\left(g_+^{-1}(x)\pa_{\beta}g_+(x)f_\g\right)
  =\frac{2}{\al^2}\delta_{\al,\g}
\een
In \cite{deBF} essentially this trace is introduced as a key object
in the explicit Wakimoto construction in those papers 
(and we see here that our $\vab$ 
is related to the matrix inverse of that). In the 
present paper we explicitly evaluate this trace (or equivalently 
$\vab$) in terms of a simple universal analytic function of the matrix $C(x)$.
A similar but somewhat more complicated expression was provided in \cite{Ito0}.
Analogous and new results will be given for all the other objects occurring:
The remaining $V$'s as well as the  $P$'s and the $Q$'s. These results
are obtained by (laboriously) working out the Gauss decompositions 
(\ref{Gauss1}) involved.

In \cite{Tay} the $V$'s are determined by an approach very similar to the
one we have employed. However, again we have carried out explicitly the
Gauss decomposition. In \cite{Tay} functions similar to
the $P$'s are given by recursion relations while functions similar to the 
$Q$'s are not discussed.

The Gauss decompositions rely on the Campbell-Baker-Hausdorff (CBH) formula
(see e.g. \cite{JR} for a proof)
\bea
 e^Ae^{tB}&=&\exp\left\{A+t\sum_{n\geq 0}\frac{B_n}{n!}(-\mbox{ad}_A)^nB
  +{\cal O}(t^2)\right\}
\label{CBH}
\eea
where the coefficients $B_n$ are the Bernoulli numbers
\bea
  B(u)&=&\frac{u}{e^u-1}=\sum_{n\geq 0}\frac{B_n}{n!}u^n\nn
  B(u)-B(-u)=-u\ &,&\ B_{2m+1}=0\ \ \ \ \ \ \ {\rm for}\ \ m\geq 1\nn
  B_0=1\ \ ,\ \ B_1&=&-\hf\ \ ,\ \ B_2=\frac{1}{6}\ \ ,\ \ B_4=-\frac{1}{30}\
   \ ,...\nn
  B^{-1}(u)&=&\frac{e^u-1}{u}=\sum_{n \geq 1}\frac{u^{n-1}}{n!}
\label{Ber}
\eea
We apply these repeatedly (infinitely many times) to the group element 
$g_+(x)e^{tj_a}$. The results are expressed in terms of the generating function
of Bernoulli numbers (\ref{Ber}) (and other even simpler analytic functions)
evaluated on the matrix $C(x)$. Since for any given Lie algebra this matrix is 
nilpotent, the formal power series all become polynomials.
The main new result of this section is then the following explicit expressions
for the polynomials $V$ and $P$ (and $Q$) in the
differential operator realization (\ref{defVP}) of the Lie algebra {\bf g}:
\bea
 \vab(x)&=&\left[B(C(x))\right]_\al^\beta\nn
 \vib(x)&=&-\left[C(x)\right]_i^\beta \nn
 \vmab(x)&=&\left[e^{-C(x)}\right]_{-\al}^\g\left[B(-C(x))\right]_\g^\beta\nn
 \paj(x)&=&\left[e^{-C(x)}\right]_{-\al}^j \nn
  Q_{-\al}^{-\beta}(x)&=&\left[e^{-C(x)}\right]_{-\al}^{-\beta}
\label{VPQ}
\eea
These matrix functions are defined in terms of universal
power series expansions, valid for any Lie algebra, but ones that truncate and
give rise to polynomials the orders of which do depend on the algebra, see
\cite{JR} for details on how the truncations work and for an alternative
explicit polynomial expression of $V_{-\al}^\beta(x)$. 

Now we introduce a differential operator \cite{FF,BMP,Ito,ATY}
which will turn out to be a building
block in the free field construction of screening operators of both the first 
and the second kinds in Section 5. It is the 
differential operator $S_\al$ (which we may construct for any root, although 
only the ones for simple roots will be used). It is defined in terms of a 
left action
\bea
 \exp\{-te_\al\}g_+(x)&=&\exp\{tS_\al(x,\pa)+{\cal O}(t^2)\}g_+(x)\nn
 S_\al(x,\pa)&=&S_\al^\beta(x)\db
\eea
It is easily seen that
\ben
 S_\al(x,\pa)=\tilde{E}_\al(-x,-\pa)
\een
so that
\ben
 S_\al^\beta(x)=-\left[B(-C(x))\right]_\al^\beta
\een
{}From the associativity property of
\ben
 e^{-se_\al}g_+(x)e^{tj_a}
\een
and the Gauss decomposition (\ref{Gauss1}) one deduces the following 
commutation relations
\bea
 \left[\tilde{E}_\al(x,\pa),S_\beta(x,\pa)\right]&=&0\nn
 \left[\tilde{H}_i(x,\pa,\Lambda),S_\beta(x,\pa)\right]&=&(\al_i^\vee,\beta)
  S_\beta(x,\pa)\nn
 \left[\tilde{F}_\al(x,\pa,\Lambda),S_\beta(x,\pa)\right]&=&
 \paj(x)(\al_j^\vee,\beta)S_\beta(x,\pa)+Q_{-\al}^{-\g}(x)(\delta_{\beta\g}
  (\beta^\vee,\Lambda)-{f_{\beta,-\g}}^\sigma S_\sigma(x,\pa))\nn  
 \left[S_\al(x,\pa),S_\beta(x,\pa)\right]&=&{f_{\al\beta}}^\g S_\g(x,\pa)
\label{Scomm}
\eea
The last commutator follows from the associativity of $e^{-se_\al}e^{-te_\beta}
g_+(x)$.
                                         
We conclude this section by listing certain 
{\em classical polynomial identities} 
(as opposed to quantum polynomial identities established in Section 4) 
needed in subsequent sections:
\bea
 (\al_i^\vee,\beta-\al)V_\al^\beta(x)&=&
  (\al_i^\vee,\g)x^\g\pa_\g V_\al^\beta(x)\nn
 (\al_i^\vee,\g+\al)V_{-\al}^\g(x)&=&
  (\al_i^\vee,\beta)x^\beta\pa_\beta V_{-\al}^\g(x)\nn
 V_\al^\g(x)\pa_\g V_\beta^\sigma(x)-V_\beta^\g(x)\pa_\g V_\al^\sigma(x)&=&
  {f_{\al\beta}}^\g V_\g^\sigma(x)\nn
 V_\al^\g(x)\pa_\g V_{-\beta}^\sigma(x)-V_{-\beta}^\g(x)\pa_\g V_\al^\sigma(x) 
 &=&{f_{\al,-\beta}}^\g V_\g^\sigma(x)+{f_{\al,-\beta}}^{-\g}V_{-\g}^\sigma(x)
  -\delta_{\al\beta}(\al^\vee,\sigma)x^\sigma\nn
 V_{-\al}^\g(x)\pa_\g V_{-\beta}^\sigma(x)-V_{-\beta}^\g(x)\pa_\g 
  V_{-\al}^\sigma(x)&=&-{f_{\al\beta}}^\g V_{-\g}^\sigma(x)\nn
 V_{\al}^\beta(x)\db\paj(x)&=&G^{ij}(\al_i^\vee,\al^\vee)\nn
 V_\al^\g(x)\pa_\g P_{-\beta}^j(x)&=&{f_{\al,-\beta}}^{-\g}P_{-\g}^j(x)\nn
 (\al_i^\vee,\beta)x^\beta\pa_\beta\paj(x)&=&(\al_i^\vee,\al)\paj(x)\nn
 V_{-\al}^\g(x)\pa_\g P_{-\beta}^j(x)-V_{-\beta}^\g(x)\pa_\g\paj(x)&=&
  -{f_{\al\beta}}^\g P_{-\g}^j(x)
\label{classV}
\eea
They are obtained directly from the fact that $E_\al,H_i$ and $F_\al$ 
constitute a differential operator realization of {\bf g}. Similarly, 
(\ref{Scomm}) gives the classical identities
\bea
 V_\al^\g(x)\pa_\g S_\beta^\sigma(x)-S_\beta^\g(x)\pa_\g V_\al^\sigma(x)&=&0\nn
 (\al_i^\vee,\beta-\al)S_\al^\beta(x)&=&(\al_i^\vee,\g)
  x^\g\pa_\g S_\al^\beta(x)\nn
 V_{-\al}^\g(x)\pa_\g S_\beta^\sigma(x)-S_\beta^\g(x)\pa_\g
  V_{-\al}^\sigma(x)&=&\paj(x)(\al_j^\vee,\beta)S_\beta^\sigma(x)
  -Q_{-\al}^{-\g}(x){f_{\beta,-\g}}^\mu S_\mu^\sigma(x)\nn
 S_\al^\g(x)\pa_\g S_\beta^\sigma(x)-S_\beta^\g(x)\pa_\g S_\al^\sigma(x)&=&
  {f_{\al\beta}}^\g S_\g^\sigma(x)\nn
 S_\beta^\g(x)\pa_\g \paj(x)&=&-Q_{-\al}^{-\beta}(x)
  (\al_i^\vee,\beta^\vee)G^{ij}
\label{classS}
\eea

\section{Wakimoto free field realizations}

The free field realization is well known to be obtained from the differential 
operator realization $\left\{\tilde{J}_a\right\}$ by the substitution
\cite{FF,GMMOS,BMP,Ito,Ku,ATY}
\ben
 \pa_\al\rightarrow\beta_\al(z)\spa x^\al\rightarrow\g^\al(z)
  \spa\Lambda_i\rightarrow\kvt\pa\varphi_i(z)
\een
and a subsequent normal ordering contribution or anomalous term, 
$F^{\mbox{{\footnotesize anom}}}_\al(\g(z),\pa\g(z))$, 
to be added to the lowering part. This term must have conformal
dimension 1, and hence is bound to be of the form
\ben
 F^{\mbox{{\footnotesize anom}}}_\al(\g(z),\pa\g(z))
  =\pa\g^{\beta}(z)F_{\al\beta}(\g(z))
\een
giving rise to the following form of the free field realization
\bea
 E_\al(z)&=&:\vab(\g(z))\beta_\beta(z):\nn
 H_i(z)&=&:\vib(\g(z))\beta_\beta(z):+\kvt\pa\varphi_i(z)\nn
 F_\al(z)&=&:\vmab(\g(z))\beta_\beta(z):+\kvt\pa\varphi_j(z)\paj(\g(z))
  +\pa\g^\beta(z)F_{\al\beta}(\g(z))\nn
 \Delta(J_a)&=&1
\label{Wakimoto}
\eea
where the normal ordering part for a {\em simple} root has been known for 
some time (\cite{Ito0})
\ben
 \pa\g^\beta(z) F_{\al_i\beta}(\g(z))=
  \pa\g^{\al_i}(z) \left(\frac{k+t}{\al_i^2}-1\right)
\een
To find the result in the general case we first derive {\em quantum 
polynomial identities} obtained by imposing the correct form of the
OPE of the form $JF$ and $TF$ (we leave out the argument $z$):
\bea
 \frac{2k}{\al^2}\delta_{\al,\beta}&=&
  -\pa_\sigma V_\al^\g\pa_\g V_{-\beta}^\sigma+V_\al^\g F_{\beta\g}
  \nn
 {f_{\al,-\beta}}^{-\g}\pa\g^\delta F_{\g\delta}&=&
  -\pa\g^\sigma\pa_\sigma\pa_\mu V_\al^\g\pa_\g V_{-\beta}^\mu+
 V_\al^\g\pa\g^\delta\pa_\g
  F_{\beta\delta} +\pa\g^\sigma\pa_\sigma V_\al^\g F_{\beta\g}\nn
 0&=&(\al_i^\vee,\sigma)\pa_\sigma V_{-\beta}^\sigma-(\al_i^\vee,\al)\g^\al
  F_{\beta\al}+tG_{ij}P_{-\beta}^j\nn
 (\al_i^\vee,\beta)\pa\g^\delta F_{\beta\delta}&=&
  (\al_i^\vee,\al)\g^\al\pa\g^\delta\pa_\al F_{\beta\delta}
  +(\al_i^\vee,\al)\pa\g^\al F_{\beta\al}\nn
 0&=&2(\rho,\al_j^\vee)P_{-\al}^j+\pa_\g V_{-\al}^\g\nn
 \pa_\g V_{-\al}^\sigma\pa_\sigma V_{-\beta}^\g&=&
  tG_{ij}P_{-\al}^iP_{-\beta}^j
  +V_{-\al}^\g F_{\beta\g}+V_{-\beta}^\g F_{\al\g}\nn
 {f_{\al\beta}}^\g F_{\g\sigma}&=&\pa_\sigma\pa_\g 
  V_{-\al}^\mu\pa_\mu V_{-\beta}^\g-V_{-\al}^\g\pa_\g 
  F_{\beta\sigma}+V_{-\beta}^\g\pa_\g F_{\al\sigma}\nn
 &-&tG_{ij}\pa_\sigma P_{-\al}^iP_{-\beta}^j
  -\pa_\sigma V_{-\al}^\g F_{\beta\g}-V_{-\beta}^\g\pa_\sigma F_{\al\g}
\label{quanpol}
\eea
from the OPE's $EF$, $HF$, $TF$ and $FF$.
Not all the identities are independent, e.g. the second to last one follows
from the last. The form of the normal ordering term is completely 
determined from the first identity, since we may introduce the inverse of 
$V_+^+\sim V_\al^\beta$.
Indeed we shall only need
\ben
 \left((V_+^+)^{-1}\right)_\al^\beta
\een 
and it follows immediately that
\bea
 (V_+^+(\g))^{-1}&=&B(C_+^+(\g))^{-1}\nn
 &=&\sum_{n\geq0}\frac{1}{(n+1)!}(C_+^+(\g))^n
\eea
Thus we have
\bea
 F_{\al\beta}(\g)&=&\frac{2k}{\al^2}
  \left((V_+^+(\g))^{-1}\right)_\beta^\al
  +\left((V_+^+(\g))^{-1}\right)_\beta^\mu\pa_\sigma V_\mu^\g(\g)\pa_\g 
  V_{-\al}^\sigma(\g)
\label{anomal}
\eea
A somewhat more involved form was given in \cite{JR}.
The present result is similar to the one in \cite{deBF}, but as before, 
in this paper we have 
provided the explicit analytic results for all the polynomials of $\g(z)$ which
enter.
\section{Screening currents}

\subsection{Screening currents of the first kind}

A screening current has conformal weight 1 and has the property
that the singular part of the OPE with an affine current is a total 
derivative. These properties ensure that integrated
screening currents (screening charges) may be inserted into correlators
without altering the conformal or affine Ward identities. This in turn makes 
them very useful in construction of correlators, see e.g. 
\cite{DF,BF,Dot,PRY1,PRY2}.
The best known screening currents \cite{FF,BMP,Ito0,Ku,ATY,deBF} 
are the following denoted screening currents of the first kind, one for each 
simple root
\bea
 s_j(z)&=&:S_{\al_j}^\al
  (\g(z))\beta_\al(z)e^{-\frac{1}{\kvt}\al_j\cdot\varphi(z)}: \nn
\al_j\cdot\var(z)&=&\frac{\al_j^2}{2}\var_j(z)
\label{sj}
\eea
In this case we find
\bea
  E_\al(z)s_j(w)&=&0\nn
  H_i(z)s_j(w)&=&0\nn
  F_\al(z)s_j(w)&=&-\frac{2t}{\al_j^2}\frac{\pa}{\pa w}\left(
    \frac{1}{z-w}Q_{-\al}^{-\al_j}(\g(w))
  :e^{-\frac{1}{\kvt}\al_j\cdot\var(w)}:\right)\nn
  T(z)s_j(w)&=&\frac{\pa}{\pa w}\left(\frac{1}{z-w}s_j(w)\right)
\eea
In our formalism the proof of these relations is a matter of direct 
verification and straightforward for $E_\al,\ H_i$ and $T$ using the 
classical polynomial identities.
In order for the OPE $F_\al(z) s_j(w)$ to be a total derivative
we find that the following two relations are sufficient conditions
\bea
 \pa_\g V_{-\al}^\beta\pa_\beta S_{\al_j}^\g&=&
  2\left(1-\frac{t}{\al_j^2}\right)
  S_{\al_j}^\beta\pa_\beta P_{-\al}^j+S_{\al_j}^\g F_{\al\g}\nn
 S_{\al_j}^\beta\pa_\beta F_{\al\sigma}&=&\pa_\g V_{-\al}^\beta
  \pa_\beta\pa_\sigma S_{\al_j}^\g-A_{ij}\pa_\sigma S_{\al_j}^\beta
  \pa_\beta P_{-\al}^i
 -\pa_\sigma S_{\al_j}^\g F_{\al\g}
\label{sufffirst}
\eea
They are easily verified for $\al$ a simple root. In the case of a non-simple
root $\al$ we have proven the conditions
(\ref{sufffirst}) by induction in addition of roots
using various classical and quantum polynomial identities. 

In \cite{FFR}\footnote{We thank E. Frenkel for pointing out this work to us.} 
screening currents of the first kind are considered and a proof of their
properties
is presented. In the recent work \cite{deBF} a more direct proof similar to the
one above is provided.

\subsection{Screening currents of the second kind}

The best known screening current of the second kind
is the one by Bershadsky and Ooguri for $SL(2)$ \cite{BO}. For non-integral
representations it involves non-integer powers of free ghost fields. 
Therefore, for some time discussions
on its interpretation were only partly successful \cite{Dot,FGPP}. 
However, in the series
of papers \cite{PRY1,PRY2} we have provided techniques based on fractional
calculus for handling such objects. Those techniques directly
generalize to the present more general situation. Screening currents of 
both kinds are necessary for being able to treat correlators of primary fields
belonging to degenerate (in particular admissible) representations 
\cite{KK,KW}.

The following expression for the screening currents of the second kind was 
written down without proof by Ito \cite{Ito0}
\bea
   \tilde{s}_j(w)&=&:\left(S_{\al_j}^{\beta}(\g(w))\beta_{\beta}(w) 
   e^{-\frac{1}{\kvt}\al_j\cdot\varphi(w)} \right)^{-\frac{2t}{\al_j^2}}:\nn
&=&:\left(S_{\al_j}^{\beta}(\g(w))\beta_{\beta}(w)\right)
^{-\frac{2t}{\al_j^2}}::e^{\sqrt{t}\var_j(w)}: 
\label{sjt1}
\eea
Here we will show that (at least in the case of $SL(N)$) 
they satisfy
\bea
  E_\al(z)\tilde{s}_j(w)&=&0\nn
  H_i(z)\tilde{s}_j(w)&=&0\nn
 F_\al(z)\tilde{s}_j(w)&=&
  -\frac{2t}{\al_j^2}\frac{\pa}{\pa w}\left(\frac{1}{z-w}:Q_{-\al}^{-\al_j}
  (\g(w))\left(S_{\al_j}^{\beta}(\g(w))\bb(w)\right)^{-\frac{2t}{\al_j^2}-1}
   e^{\kvt\var_j(w)}: \right) \nn
   T(z)\tilde{s}_j(w)&=&\frac{\pa}{\pa w}\left(\frac{1}{z-w}\tilde{s}_j(w)
  \right)
\label{sjt2}
\eea
We employ the techniques discussed in \cite{PRY1,PRY2} for
how to perform contractions involving ghost fields raised to non-integer
powers. Such techniques are necessary in the generic case where 
$-\frac{2t}{\al_j^2}$ is not integer. 

In the case of $SL(N)$ where a simple 
root (here $\al_j$) appears at most once in the decomposition of a positive
root, it is straightforward to check that 
$H_i(z)\tilde{s}_j(w)=0$ and $\Delta(\tilde{s}_j)=1$.

Let us introduce the shorthand notation
\ben
 S_j^u(z)=:\left(S_{\al_j}^\beta(\g(z))\bb(z)\right)^u:
\een
and consider
\bea
 E_\al(z)S_j^u(w)&=&\sum_{l\geq1}\frac{1}{(z-w)^l}(-1)^l\binomial{u}{l}
  \left(:\pa_{\g_1}...\pa_{\g_l}\vab\bb(z)S_{\al_j}^{\g_1}...S_{\al_j}^{\g_l}
  S_j^{u-l}:\right.\nn
  &-&l\left.:\pa_{\g_1}...\pa_{\g_{l-1}}\vab(z)S_{\al_j}^{\g_1}...S_{\al_j}^{
  \g_{l-1}}
  \db S_{\al_j}^{\g_l}\beta_{\g_l}S_j^{u-l}:\right)\nn
 &+&\sum_{l\geq 2}\frac{(-1)^{l-1}}{(z-w)^l}\binomial{u}{l-1}:\pa_{\g_1}...
  \pa_{\g_{l-1}}V_\al^\beta(z)\pa_\beta(S_{\al_j}^{\g_1}...S_{\al_j}^{\g_{l-1}}
  )S_j^{u-l+1}:
\eea
Here and in the following equations we have suppressed the argument $\g(z)$ of
${V_\al}^\beta$ and other fields. We use the explicit expressions for
$V_\al^\beta$ and $S_{\al_j}^\g$ and find that all terms for $l>1$ vanish.
Hence, in that case we get
\bea
 E_\al(z)S_j^u(w)&=&\frac{u}{z-w}:\left(V_\al^\g\pa_\g S_{\al_j}^\beta-
  S_{\al_j}^\g\pa_\g V_\al^\beta\right)\beta_\beta S_j^{u-1}:(w)\nn
 &=&0
\eea
In a similar way, we have worked out the 
OPE $F_\al(z)\tilde{s}_j(w)$ and in the case of
$SL(N)$ it reduces to ($u=-\frac{2t}{\al_j^2}$)
\bea
 F_\al(z)\tilde{s}_j(w)&=&\frac{1}{(z-w)^2}\left(-u:\pa_\g V_{-\al}^\beta(z)
  \pa_\beta S_{\al_j}^\g S_j^{u-1}:\right.\nn
 &+&\frac{u(u-1)}{2}:\pa_{\g_1}
  \pa_{\g_2}V_{-\al}^\beta\beta_\beta(z)
  S_{\al_j}^{\g_1}S_{\al_j}^{\g_2}S_j^{u-2}:\nn
 &-&u(u-1):\pa_{\g_1}V_{-\al}^\beta(z)S_{\al_j}^{\g_1}\pa_\beta 
  S_{\al_j}^{\g_2}\beta_{\g_2}S_j^{u-2}:\nn
 &-&\left. utG_{ij}:\pa_\g P_{-\al}^i(z)S_{\al_j}^\g S_j^{u-1}:+u:F_{\al\sigma}
  (z)S_{\al_j}^\sigma S_j^{u-1}:\right):e^{\sqrt{t}\var_j(w)}:\nn
 &+&\frac{1}{z-w}\left(u:(V_{-\al}^\g\pa_\g S_{\al_j}^\beta-S_{\al_j}^\g\pa_\g
  V_{-\al}^\beta)\beta_\beta S_j^{u-1}::e^{\sqrt{t}\var_j(w)}:\right.\nn
 &-&u\kvt:\pa_\g P_{-\al}^iS_{\al_j}^\g S_j^{u-1}\pa\var_i
  e^{\sqrt{t}\var_j(w)}:\nn
 &+&\left.tG_{ij}:P_{-\al}^iS_j^u::e^{\sqrt{t}\var_j(w)}:
  -u:\pa\g^\sigma\pa_\g F_{\al\sigma}
  S_{\al_j}^\g S_j^{u-1}::e^{\sqrt{t}\var_j(w)}:\right)
\eea
A comparison with (\ref{sjt2}) 
yields the following consistency condition
\bea
 &&(u-1)\left(\hf\pa_\sigma\pa_{\g_1}\pa_{\g_2}V_{-\al}^\beta S_{\al_j}^{\g_1}
  S_{\al_j}^{\g_2}-\pa_\sigma\pa_{\g_1}V_{-\al}^{\g_2}S_{\al_j}^{\g_1}
  \pa_{\g_2}S_{\al_j}^\beta\right)\nn
 &+&\left(-\pa_\sigma\pa_\g V_{-\al}^\mu\pa_\mu S_{\al_j}^\g
  +S_{\al_j}^\g\pa_\sigma F_{\al\g}
  -tG_{ij}S_{\al_j}^\g\pa_\g\pa_\sigma P_{-\al}^i -S_{\al_j}^\g\pa_\g 
  F_{\al\sigma}\right)S_{\al_j}^\beta\nn
 &=&(u-1)Q_{-\al}^{-\al_j}\pa_\sigma S_{\al_j}^\beta
  +\pa_\sigma Q_{-\al}^{-\al_j}S_{\al_j}^\beta
\label{suffsec}
\eea
besides more trivial relations such as
\bea
 S_{\al_j}^\g\pa_\g P_{-\al}^i&=&-\delta_j^iQ_{-\al}^{-\al_j}\nn
 \hf S_{\al_j}^{\g_1}S_{\al_j}^{\g_2}
  \pa_{\g_1}\pa_{\g_2}V_{-\al}^\beta&=&Q_{-\al}^{-\al_j}S_{\al_j}^\beta\nn
 S_{\al_j}^{\g_1}\pa_{\g_1}V_{-\al}^{\g_2}\pa_{\g_2}S_{\al_j}^\beta&=&0
\eea
which are easily seen to be satisfied. One can verify the less trivial part
(\ref{suffsec}) using the polynomial identities
together with the consistency conditions (\ref{sufffirst}). 
Hence, we conclude that in the case of $SL(N)$ the screening currents 
of the second kind (\ref{sjt2}) exist. As already mentioned it seems natural
that the expression (\ref{sjt2}) should hold for all simple groups; we
refer to \cite{JR2} (and \cite{JR}) for further details. 
In \cite{JR2} a quantum group
structure based on both kinds of screening currents will also be discussed
(see also the presentation in \cite{JR}), along the lines of Gomez and 
Sierra \cite{GS}.

\section{Primary fields}

The final new result reported in this paper is the explicit construction in
this section of primary fields for arbitrary representations, integral or 
non-integral (for integral representations, see also \cite{FF}). 
We find it particularly convenient to replace the traditional 
multiplet of primary fields (which generically would be infinite for 
non-integrable representations) by a generating function for that, namely 
the primary field $\phi_\La(w,x)$ which must satisfy
\bea
 J_a(z)\phi_\La(w,x)&=&\frac{-J_a(x,\pa,\La)}{z-w}\phi_\La(w,x)\nn
 T(z)\phi_\La(w,x)&=&\frac{\Delta(\phi_\La)}{(z-w)^2}\phi_\La(w,x)
  +\frac{1}{z-w}\pa\phi_\La(w,x)
\label{primdef}
\eea
Here $J_a(z)$ are the affine currents, whereas $J_a(x,\pa,\Lambda)$ are the 
differential operator realizations 
Eqs. (\ref{ketdef}), (\ref{tilde}), (\ref{defVP}) and (\ref{VPQ}).
For the simplest case of affine $SL(2)$ the result is known (\cite{FGPP,PRY1}).
In that case we have only one positive root, one $x$, one ghost pair
$(\beta(z),\g(z))$ and one scalar field $\varphi(z)$ while $\La$ is
given by the spin $j$ ($2j=\La_1$ is non-integral in general). The 
result is
\ben
 \phi_j(w,x)=(1+x \g(w))^{2j} :e^{\frac{2j}{\kvt}\varphi(w)}:
\een
In this section we show how to generalize this sort of expression to an
arbitrary Lie algebra, with particularly explicit prescriptions in the
case of affine $SL(N)$.
We shall find the result in the form
\bea
 \phi_\La(w,x)&=&\phi_\La'(\g(w),x)V_\Lambda(w) \nn
 V_\Lambda(w)&=&:e^{\frac{1}{\kvt}\Lambda\cdot\var(z)}:\nn
\phi_\La'(\g(w),0)&=&1
\label{primans}
\eea
Indeed such a field is conformally primary and has conformal dimension
$\Delta(\phi_\La)=\frac{1}{2t}(\Lambda,\Lambda+2\rho)$. In order to comply with
(\ref{primdef}) for $J_a=H_i$, it seems very plausible that $\phi_\La'$ 
must be symmetric in $\g(w)$ and $x$. Below we shall show this by explicit 
construction. Due to the fact that the anomalous or normal ordering part of 
$F_\al(z)$ does not give singular contributions
when contracting with $\phi_\La'$, it is then 
sufficient to consider OPE's with $E_\al$. The point is that the two OPE's 
$E_\al(z)\phi_\La(w,x)$ and $F_\al(z)\phi_\La(w,x)$ are obtained from one 
another by interchanging $x$ and $\g(w)$. Because of the above symmetry it
is enough to verify one of them. We therefore obtain the following 
sufficient conditions on $\phi_\La'(\g(w),x)$, one for each $\al>0$
\ben
 V_\al^\beta(\g)\pa_{\g^\beta}\phi_\La'=V_{-\al}^\beta(x)\pa_{x^\beta}\phi_\La'
  +\La_jP_{-\al}^j(x)\phi_\La'
\label{primsuff}
\een
Further, one can use the classical polynomial 
identities (\ref{classV}) to demonstrate that if $\phi_\La$ is a
primary field wrt $E_\al$ and $E_\beta$ then it is a primary field wrt
${f_{\al\beta}}^\g E_\g$. Effectively, this amounts to prove (\ref{primsuff})
for a sum $\al=\beta+\g$ of two roots under the assumption that it is
satisfied for both $\al=\beta$ and $\al=\g$.
This means that there are only $r$ sufficient
conditions a primary field (\ref{primans}) must satisfy
\ben
 V_{\al_i}^\beta(\g)\pa_{\g^\beta}\phi_\La'=
  V_{-\al_i}^\beta(x)\pa_{x^\beta}\phi_\La'
  +\Lambda_ix^{\al_i}\phi_\La'
\label{primeq}
\een 
It is very hard to solve this set of partial differential equations directly.
However, we have found an alternative way to obtain the primary fields.
The construction goes as follows.

First we directly construct primary fields for each fundamental representation
$M_{\La^{(k)}}$. Such representation spaces are finite dimensional modules 
and $\phi'_{\La^{(k)}}(\g(w),x)$ will be polynomial in $\g(w)$ and $x$.
Then finally, for a general representation with highest weight 
$\La=\La_k\La^{(k)}$
we use (\ref{primeq}) to immediately obtain that\footnote{In \cite{JR} a 
discussion of this point is partly incorrect.}
\ben
 \phi'_{\La}(\g(w),x)=\prod_{k=1}^r{[}\phi'_{\La^{(k)}}(\g(w),x){]}^{\La_k}
\label{ansatz}
\een
We emphasize here that the dynkin labels, $\La_k$, may be non-integers as is
required for degenerate (including admissible) representations.
We proceed to explain how to construct the building blocks
\ben
 \phi'_{\La^{(k)}}(\g(w),x)
\een

The strategy goes as follows. First we concentrate on the case $w=0$ where
the object reduces to
\ben
 \phi'_{\La^{(k)}}(\g_0,x)
\label{phimgo}
\een
when acting on the highest weight state $\ket{\La^{(k)}}$.
$\g_0$ is the zero mode in the mode expansion
\ben
 \g(w)=\sum_n \g_n w^{-n}\spa \beta(w)=\sum_n\beta_nw^{-n-1}
\een
Conformal covariance requires $\phi'_{\La^{(k)}}(\g(w),x)$ to be obtained just
by replacing $\g_0$ by $\g(w)$. The function (\ref{phimgo}) 
in turn is obtained from
\ben
 \ket{\La^{(k)},x}=g_-(x)\ket{\La^{(k)}}=\phi'_{\La^{(k)}}(\g_0,x)
  \ket{\La^{(k)}}
\een
Indeed, it is an immediate consequence of the formalism, that the primary
field constructed this way will satisfy the OPE (\ref{primdef}). The
construction is now simply achieved by expanding the state
$\ket{\La^{(k)},x}$ on an appropriate basis which is convenient to obtain
using the free field realization.

Let the orthonormal basis elements in the $k$'th fundamental highest 
weight module $M_{\La^{(k)}}$ be 
denoted $\{\ket{b,\La^{(k)}}\}$ such that the identity operator may be written
\ben
 I=\sum_b\ket{b,\La^{(k)}}\bra{b,\La^{(k)}}
\label{identity}
\een
The state $\ket{\La^{(k)},x}$ may then be written
\ben
 \ket{\La^{(k)},x}=\sum_b\ket{b,\La^{(k)}}\bra{b,\La^{(k)}}\La^{(k)},x\rangle
\een
One of the basis vectors will always be taken to be the highest weight 
vector $\ket{\La^{(k)}}$ itself. 

Now concentrate on one particular basis vector. It will be of the form
\ben
 \ket{b,\La^{(k)}}=f_{\beta_1^{(b)}}...f_{\beta_{n(b)}^{(b)}}\ket{\La^{(k)}}
\een
and the expansion of $\ket{\La^{(k)},x}$ will be
\ben
 \ket{\La^{(k)},x}=\sum_b f_{\beta_1^{(b)}}...f_{\beta_{n(b)}^{(b)}}
  \ket{\La^{(k)}}
 \bra{\La^{(k)}}e_{\beta_{n(b)}^{(b)}}...e_{\beta_1^{(b)}}\ket{\La^{(k)},x}
\label{ILax}
\een
For each term in the sum we treat the two factors differently. First consider 
the second factor. We may use the differential operator realizations to write
\bea
 \bra{\La^{(k)}}e_{\beta_{n(b)}^{(b)}}...e_{\beta_1^{(b)}}\ket{\La^{(k)},x}
 &=&(-1)^{n(b)}E_{\beta_1^{(b)}}(x,\pa,\La^{(k)})
  ...E_{\beta_{n(b)}^{(b)}}
  (x,\pa,\La^{(k)})\bra{\La^{(k)}}\La^{(k)},x\rangle\nn
 &=&b(x,\La^{(k)})
\eea
where
\ben
 b(x,\La^{(k)})=\left[V^{\g_1}_{-\beta_1^{(b)}}(x)\pa_{\g_1}+
  P^k_{-\beta_1^{(b)}}(x)\right]...
  \left[V^{\g_{n(b)-1}}_{-\beta_{n(b)-1}^{(b)}}(x)
  \pa_{\g_{n(b)-1}}+ P^k_{-\beta_{n(b)-1}^{(b)}}(x)\right]
  P^k_{-\beta_{n(b)}^{(b)}}(x)
\label{primpol}
\een
In the last step we used that clearly
\ben
 \bra{\La^{(k)}}\La^{(k)},x\rangle \equiv 1
\een                             
In the first factor in (\ref{ILax}) 
\ben
  f_{\beta_1^{(b)}}...f_{\beta_{n(b)}^{(b)}}\ket{\La^{(k)}}
\een
we use the free field realizations. The state $\ket{\La^{(k)}}$ is a vacuum for
the $\beta,\g$ system, so it is annihilated by $\g_n,\ n\geq 1$ and by 
$\beta_n,\ n\geq 0$. The $f_\beta$'s are the zero modes of the affine currents.
It follows that only $\g_0$'s and $\beta_0$'s need be considered. Also 
the normal ordering term will not give a contribution, and we obtain
\bea
 f_{\beta_1^{(b)}}...f_{\beta_{n(b)}^{(b)}}\ket{\La^{(k)}}
 &=&\left[V^{\g_1}_{-\beta_1^{(b)}}(\g_0)\beta_{\g_1,0}+
 P^k_{-\beta_1^{(b)}}(\g_0)\right]...\nn
 &\cdot&\left[V^{\g_{n(b)-1}}_{-\beta_{n(b)-1}^{(b)}}(\g_0)
  \beta_{\g_{n(b)-1},0}+ P^k_{-\beta_{n(b)-1}^{(b)}}(\g_0)\right]
  P^k_{-\beta_{n(b)}^{(b)}}(\g_0)\ket{\La^{(k)}}\nn
 &=&b(\g_0,\La^{(k)})\ket{\La^{(k)}}
\eea
By the remarks above this completes the construction in general. Explicit 
expressions for the $V$'s and the $P$'s have already been provided.

It remains to account in detail for how to obtain a convenient basis for the 
fundamental representations. This part will depend on the algebra. Here
for completeness and illustration we indicate the construction for $SL(N)$
where the fundamental representations are conveniently realized in terms of 
$N$ fermionic creation and annihilation operators 
\ben
 q_i^\dagger,q_i, \ \ i=1,...,N
\een
An orthonormal basis in the $k$'th fundamental representation is provided 
by the sets of states where $k$ fermionic creation operators act on the Fermi
vacuum, giving dimension 
\ben
 \binomial{N}{k}
\een
The $r=N-1$ dimensional root and weight space may be represented as the
$N-1$ dimensional hyperplane in $N$ dimensional euclidean space 
\ben
 \{\sum_{j=1}^N y^j\vec{e}_j\ |\ \sum_j y^j=0\}
\een
where $\{\vec{e}_j\}$ is an orthonormal basis for the $N$ dimensional space.
The roots of $SL(N)$ are of the form 
\ben
 \al_{ij}=\vec{e}_i-\vec{e}_j
\een
We may take
\bea
 e_{ij}&\equiv&e_{\al_{ij}}=q_i^\dagger q_j\spa i<j\nn
 f_{ij}&\equiv&f_{\al_{ij}}=q_j^\dagger q_i\spa i<j
\eea
The highest weight vector is
\ben
 \ket{\La^{(k)}}=q_1^\dagger q_2^\dagger ... q_k^\dagger\ket{0}
\een
A basis with a minimal set of lowering operators is then easily seen to be
the set
\bea
 &&\ket{\La^{(k)}}\nn
 &&f_{ij}\ket{\La^{(k)}}, \ \ i\leq k<j\leq N \nn
 &&\vdots  \nn
 &&f_{i_1j_1}...f_{i_pj_p}\ket{\La^{(k)}}, \ \ 
  i_1<...<i_p\leq k < j_1 <...<j_p\leq N       \nn
 &&\vdots
\label{basis1}
\eea
where altogether $p=0,1,...,N-k$.

The primary field for the $k$'th fundamental representation is then of the form
\bea
 \phi_{\La^{(k)}}(z,x)&=&\phi'_{\La^{(k)}}(\g(z),x)V_{\La^{(k)}}(z)\nn
 \phi'_{\La^{(k)}}(\g(z),x)&=&\sum_{p=0}^{N-k}
  \sum_{i_1<...<i_p\leq k<j_1<...<j_p\leq N}\nn
 &\cdot&b_p(\{i_l\},\{j_l\},\g(z),\La^{(k)})
  b_p(\{i_l\},\{j_l\},x,\La^{(k)})\nn
 b_0(x,\La^{(k)})&=&1\nn
 b_p(\{i_l\},\{j_l\},x,\La^{(k)})&=&(-1)^pE_{\al_{i_1j_1}}(x,\pa,\La^{(k)})...
  E_{\al_{i_pj_p}}(x,\pa,\La^{(k)})1
\label{kthslN}
\eea

Actually in this particular case of $SL(N)$ (and possibly with a suitable 
generalization, for more general groups), an even more explicit realization is
possible, one not involving derivatives. Indeed the basis for the $k$'th 
fundamental representation (\ref{basis1}) 
may be equivalently obtained as the set
\ben
\ket{b;I(k)}=q^\dagger_{i_1}... q^\dagger_{i_k}\ket{0}
\label{basis2}
\een
Here we have denoted by $I(k)$ the subset $\{i_1,...,i_k\}$ of the set 
$\{1,2,...,N\}$ and  we shall denote by $M(N,k)$ the set of all these subsets,
so that
\ben
|M(N,k)|=\binomial{N}{k}
\een
Now we may evaluate
\bea
 \ket{\La^{(k)},x}&=&e^{f(x)}q^\dagger_1...q^\dagger_k\ket{0}\nn
 &=&{\left (e^{F^{(N)}(x)}\right )_1}^{j_1} ...
  {\left (e^{F^{(N)}(x)}\right )_k}^{j_k} 
  q^\dagger_{j_1}...q^\dagger_{j_k}\ket{0}
\label{statex}
\eea
where $F^{(N)}(x)$ is the matrix representation of $f(x)$ in the 
$N$-dimensional defining representation of $SL(N)$:
\bea
{F^{(N)}(x)_i}^j&=&x^{ij}, \ \ \ i<j\nn
{F^{(N)}(x)_i}^j&=&0, \ \ \ i \geq j
\eea
We have to evaluate the overlap between the state (\ref{statex}) and the basis 
vector in (\ref{basis2}). The result is the well known determinant. Namely,
denote by
\ben
e^{F^{(N)}(x)}(I(k))
\een
the $k\times k$ matrix obtained from the $N\times N$ matrix $e^{F^{(N)}(x)}$
by using the first $k$ rows and the $k$ columns given by the set $I(k)$.
Then the sought overlap (up to a sign which will be irrelevant) is the 
determinant of that reduced $k\times k$ matrix. From the discussion above we
know that in the primary field, the polynomial in $x$ thus obtained will be 
multiplied by exactly the same polynomial in $\gamma(z)$ 
(hence the irrelevance of the above sign). Thus we have the
simplified version of Eq. (\ref{kthslN})
\ben
\phi'_{\La^{(k)}}(\gamma(z),x)=\sum_{I(k)\in M(N,k)}
\det\left (e^{F^{(N)}(x)}(I(k))\right )
\det\left (e^{F^{(N)}(\gamma(z))}(I(k))\right )
\label{kthslN2}
\een
A similar expression is obtained when states are represented in terms
of fermion {\em annihilation} operators acting on the ``filled Dirac sea''
state
\ben
 \ket{\overline{0}}=q_1^\dagger...q_N^\dagger\ket{0}
\een
The new form is similar to Eq. (\ref{kthslN2}) but $k$ is replaced by
$N-k$ and $F^{(N)}$ by $-F^{(N)}$, and now we have to use the {\em last}
$N-k$ rows. This second form is the most convenient one for $k\geq N/2$.

It is not difficult to check that the two expressions 
Eqs. (\ref{kthslN}) and (\ref{kthslN2}) (of course) agree with each other 
in the cases $k=1$ and $k=r=N-1$. Thus Eq. (\ref{kthslN}) gives
\bea
\phi'_{\La^{(1)}}(\g(z),x)&=&1+\sum_{j=2}^NP^1_{-\al_{1j}}(\g(z))
P^1_{-\al_{1j}}(x)\nn
\phi'_{\La^{(N-1)}}(\g(z),x)&=&1+\sum_{i=1}^{N-1}P^{N-1}_{-\al_{iN}}(\g(z))
P^{N-1}_{-\al_{iN}}(x)
\eea
As an illustration, we get in the case of $SL(3)$ 
($M_{\La^{(1)}}=\{\mbox{\bf 3}\}, \ M_{\La^{(2)}}=\{\overline{\mbox{\bf 3}}\}$)
\bea
\phi'_{\La^{(1)}}(\gamma(z),x)&=&1+\g^{12}(z)x^{12}+
(\g^{13}(z)+\hf\g^{12}(z)\g^{23}(z))(x^{13}+\hf x^{12}x^{23})\nn
\phi'_{\La^{(2)}}(\gamma(z),x)&=&1+\g^{23}(z)x^{23}+
(\g^{13}(z)-\hf\g^{12}(z)\g^{23}(z))(x^{13}-\hf x^{12}x^{23})
\eea
The last two results were obtained already in \cite{JR}.

We notice that with the two sets of screening operators constructed
in Section 5, in principle we shall be able to use the standard free field 
techniques to provide integral representations 
of correlators of primary fields with weights given by
non-integer Dynkin labels of the form
\bea
 \La_k&=&A_{ki}r^i-G_{ki}s^i t\nn
&=&\hat{r}_k-\hat{s}_k\frac{\theta^2}{\al_k^2}\hat{t} \nn
\hat{t}&=&\frac{2}{\theta^2}t = k^\vee + h^\vee
\label{dynkinlabel}
\eea
where $i,k =1,...,r$ and $r^i,s^i,\hat{r}_k,\hat{s}_k$ are integers
corresponding to degenerate representations \cite{KK}.
For $\hat{t}$ rational of the form $\hat{t}=\frac{p}{q}$ with $p$ and $q$ 
co-prime, this corresponds to admissible representations \cite{KW}.

In our previous work on $SL(2)$ \cite{PRY1,PRY2} we used a 
notation different from the one employed in this paper. 
The correspondence is the following, where
``hats'' refer to our old notation
\bea
 \hat{J}^+=E\spa\hat{J}^3&=&\hf H\spa\hat{J}^-=F\nn
 \hat{k}=k^\vee\spa \hat{t}&=&\frac{2}{\theta^2}t\spa 2\hat{j}=\La_1\nn
 \hat{\var}&=&-\sqrt{\frac{\theta^2}4}\var_1
\eea
and where $G^{11}=\frac{\theta_2}{4}$, such that $\D(\phi)=\frac{\hat{j}
(\hat{j}+1)}{\hat{t}}$. Furthermore, there are additional phases on
the screening currents. 

\section{Conclusions}
In this paper we have provided missing ingredients needed in order to use
free field realizations of affine algebras for setting up integral 
representations of conformal (chiral)
blocks for arbitrary degenerate representations
\cite{KK} generalizing the famous treatments for minimal models \cite{DF} and
the more recent ones for $SL(2)$ \cite{PRY1,PRY2}. Our new results are
(i) very explicit and universal formulas for the free field realizations of
currents, Eqs. (\ref{Wakimoto}), (\ref{anomal}), (\ref{VPQ}), (\ref{Ber}) and
(\ref{cadj}); (ii) a proof of the properties of the screening currents of 
the second kind Eq. (\ref{sjt1}) \cite{Ito0}, 
at least for affine $SL(N)$ based on the 
screening currents of the first kind Eq. (\ref{sj});
(iii) free field realizations for full multiplets of primary fields
using the triangular parameters, and valid for arbitrary weights (integral and
non-integral), Eqs. (\ref{primans}), (\ref{ansatz}) and (\ref{primpol}). 
In particular we now have ingredients for building correlators for degenerate 
(and admissible) representations with weights obtained from 
Eq. (\ref{dynkinlabel}).
The realization is particularly explicit for $SL(N)$, 
Eqs. (\ref{kthslN}) and (\ref{kthslN2}).\\[.8 cm] 
{\bf Acknowledgement}\\[.2cm]
We thank L. Feh\'er for enlightening discussions and explanations on Ref. 
\cite{deBF}. We further thank E. Frenkel for very illuminating correspondence.
This work was carried out in the framework of the project ``Gauge theories,
applied supersymmetry and quantum gravity", contract SC1-CT92-0789 of the 
European Economic Community.


\begin{thebibliography}{99}

\bibitem{Wak} M. Wakimoto, \CMP{104} (1986) 60
\bibitem{FF} B.L. Feigin and E.V. Frenkel, Usp. Mat. Nauk. {\bf 43} (1988) 227 
(in Russian), Russ. Math. Surv. {\bf 43} (1989) 221;\\
 B.L. Feigin and E.V. Frenkel, \CMP{128} (1990) 161;\\
 B.L. Feigin and E.V. Frenkel,  Lett. Math. Phys. {\bf 19} (1990) 307;\\
 B.L. Feigin and E.V. Frenkel, in {\em Physics and Mathematics of Strings},
 Eds. L. Brink {\em et al.} (World Scientific, 1990); \\
 E. Frenkel, {\em Free field realizations in representation theory and 
 conformal field theory}, preprint hep-th/9408109 
\bibitem{GMMOS} A. Morozov, JETP Lett. {\bf 49} (1989) 345;\\
 A. Gerasimov, A. Marshakov, A. Morozov, M. Olshanetskii, and S. Shatashvili, 
 Int. J. Mod. Phys. {\bf A 5} (1990) 2495
\bibitem{BF} D. Bernard and G. Felder, \CMP{127} (1990) 145
\bibitem{BMP} P. Bouwknegt, J. McCarthy and K. Pilch, \PL{B 234} (1990) 297;\\
 P. Bouwknegt, J. McCarthy and K. Pilch, \CMP{131} (1990) 125;\\
 P. Bouwknegt, J. McCarthy and K. Pilch, Prog. Theor. Phys. Suppl. {\bf 102}
 (1990) 67;\\
  P. Bouwknegt, J. McCarthy and K. Pilch in {\em Strings and symmetries 1991,}
 eds. N. Berkovits {\em et al.}, (World Scientific, Singapore, 1992)
\bibitem{KOS} M. Kuwahara and H. Suzuki, \PL{B 235} (1990) 52;\\
 M. Kuwahara, N. Ohta and H. Suzuki, \PL{B 235} (1990) 57;\\
 M. Kuwahara, N. Ohta and H. Suzuki, \NP{B 340} (1990) 448;\\
 N. Ohta and H. Suzuki, \NP{B 332} (1990) 146
\bibitem{Ito0}K. Ito, \PL{B 252} (1990) 69
\bibitem{Ito} 
 K. Ito and Y. Kazama, Mod. Phys. Lett. {\bf A 5} (1990) 215;\\
 K. Ito and S. Komata, Mod. Phys. Lett. {\bf A 6} (1991) 581
\bibitem{Dot} Vl.S. Dotsenko, \NP{B 338} (1990) 747;\\
 Vl.S. Dotsenko, \NP{B 358} (1991) 547
\bibitem{Ku} G. Kuroki, \CMP{142} (1991) 511
\bibitem{ATY} H. Awata, A. Tsuchiya and Y. Yamada, \NP{B 365} (1991) 680;
 H. Awata, Prog. Theor. Phys. Suppl. {\bf 110} (1992) 303
\bibitem{Tay} W. Taylor IV, LBL-34507, hep-th/9310040, Ph.D. thesis
\bibitem{deBF} J. de Boer and L. Feh\'er, Mod. Phys. Lett. {\bf A 11} (1996) 
 1999;\\
 J. de Boer and L. Feh\'er, {\em Wakimoto realizations of current algebras: an 
 explicit construction},  
 LBNL-39562, UCB-PTH-96/49, BONN-TH-96/16, hep-th/9611083, preprint
\bibitem{DF} Vl.S. Dotsenko and V.A. Fateev, \NP{B 240}{[}FS12{]} (1984) 312;\\
 Vl.S. Dotsenko and V.A. Fateev, \NP{B 251}{[}FS13{]} (1985) 691
\bibitem{FZ} V.A. Fateev and A.B. Zamolodchikov, \SJNP{43} (1986) 657
\bibitem{FGPP} P. Furlan, A.Ch. Ganchev, R. Paunov and V.B. Petkova,
 \PL{B 267} (1991) 63;\\
 P. Furlan, A.Ch. Ganchev, R. Paunov and V.B. Petkova,
 \NP{B 394} (1993) 665;\\
 A.Ch. Ganchev and V.B. Petkova, \PL{B 293} (1992) 56
\bibitem{An} O. Andreev, Int. J. Mod. Phys. {\bf A 10} (1995) 3221;\\
 O. Andreev, \PL{B 363} (1995) 166
\bibitem{PRY1} J.L. Petersen, J. Rasmussen and M. Yu, 
 \NP{B 457} (1995) 309;\\
 J.L. Petersen, J. Rasmussen and M. Yu, \NP{B 457} (1995) 343;\\
 J.L. Petersen, J. Rasmussen and M. Yu, in 
 {\em Proceedings of the workshop Gauge Theories, Applied Supersymmetry and
 Quantum Gravity, Leuven July 1995, Leuven Notes in Mathematical and 
 Theoretical Physics, Vol. 6}, Eds. B. de Wit {\em et al} (Leuven 1996)
 hep-th/9510059;\\
 J.L. Petersen, J. Rasmussen and M. Yu, Nucl. Phys. {\bf B} (Proc. Suppl.)
 {\bf 49} (1996) 27
\bibitem{PRY2} J.L. Petersen, J. Rasmussen and M. Yu,
\NP{B 481} (1996) 577;\\
 J.L. Petersen, J. Rasmussen and M. Yu, in {\em Proceedings of Inauguration
 Conference of the Asia Pacific Center for Theoretical Physics (APCTP),
 Seoul, Korea, 1996}
\bibitem{KK} V.G. Kac and D.A. Kazhdan, Adv. Math. {\bf 34} (1979) 97
\bibitem{KW} V.G. Kac and M. Wakimoto, Proc. Natl. Acad. Sci. 
 USA {\bf 85} (1988) 4956;\\
 V.G. Kac, and D.A. Kazhdan, Adv. Ser. Math. Phys., Vol. 7 (World Scientific, 
 1989), p. 138
\bibitem{HY} H.-L. Hu and M. Yu, \PL{B 289} (1992) 302;\\
 H.-L. Hu and M. Yu, \NP{B 391} (1993) 389
\bibitem{AGSY} O. Aharony, O. Ganor, J. Sonnenschein and S. Yankielowicz,
 \NP{B 399} (1993) 527
\bibitem{Sch} J. Schnittger, \NP{B 471} (1996) 521
\bibitem{JR2} J. Rasmussen, {\em Screening Currents of Both Kinds in 2D
 Current Algebra, and a Quantum Group Structure}, preprint in preparation
\bibitem{JR} J. Rasmussen, hep-th/9610167, Ph.D. thesis 
  (The Niels Bohr Institute) 
\bibitem{BO} M. Bershadsky and H. Ooguri, \CMP{126} (1989) 49
\bibitem{Kos} B. Kostant, Lect. Notes Math. {\bf 466} (1974) 101
\bibitem{FFR} B.L. Feigin, E.V. Frenkel and N. Reshetikhin, 
\CMP{166} (1994) 27
\bibitem{GS} C. Gomez and G. Sierra, \PL{B 240} (1990) 149;\\
 C. Gomez and G. Sierra, \NP{B 352} (1991) 791;\\
 C. Ramirez, H. Ruegg and M. Ruiz-Altaba, \PL{B 247} (1990)
 499;\\
 C. Ramirez, H. Ruegg and M. Ruiz-Altaba, \NP{B 364} (1991) 195;\\
 C. Gomez and G. Sierra, in {\em Proc. of the Trieste 1990
 Summer School} (World Scientific, 1991);\\
 C. Gomez, M. Ruiz-Altaba and G. Sierra, {\em Quantum Groups in Two-dimensional
 Physics} (Cambridge Univ. Press, 1996)

\end{thebibliography}
\end{document}